\documentclass{aa}

\usepackage{psfig}

\def\ros{{\sl ROSAT}}
\def\etal{{et\,al.}}

\def\msun{M$_{\odot}$}

\def\it{\sl}
\def\degs{\ifmmode ^{\circ}\else$^{\circ}$\fi}
\def\amin{\ifmmode ^{\prime}\else$^{\prime}$\fi}
\def\asec{\ifmmode ^{\prime\prime}\else$^{\prime\prime}$\fi}
\def\fd{\hbox{$.\!\!^{\rm d}$}}            
\def\fss{\hbox{$.\!\!^{\rm s}$}}        
\def\h{$^{\rm h}$}\def\m{$^{\rm m}$}

\newbox\grsign \setbox\grsign=\hbox{$>$}
\newdimen\grdimen \grdimen=\ht\grsign
\newbox\laxbox \newbox\gaxbox
\setbox\gaxbox=\hbox{\raise.5ex\hbox{$>$}\llap
     {\lower.5ex\hbox{$\sim$}}}\ht1=\grdimen\dp1=0pt
\setbox\laxbox=\hbox{\raise.5ex\hbox{$<$}\llap
     {\lower.5ex\hbox{$\sim$}}}\ht2=\grdimen\dp2=0pt
\def\gax{\mathrel{\copy\gaxbox}}
\def\lax{\mathrel{\copy\laxbox}}
\font\pr=cmr7

\def\II{{\pr II}}

\def\V{{\pr V}}
\def\VI{{\pr VI}}

\def\all4{V Sge, WX Cen, V617 Sge and HD 104994}
\def\0513{RX\,J0513.9--6951}

\begin{document}

\voffset=15mm

   \thesaurus{06         
              (02.01.2;  
               08.02.2;  
               08.09.2;  
               08.14.2;  
               13.25.5)} 

   \title{On the X-ray properties of V~Sge and its relation to
   the supersoft X-ray binaries}

   \author{Jochen~Greiner\inst{1} \and Andr\'e~van~Teeseling\inst{2}}

   \offprints{J. Greiner, jgreiner@aip.de}

   \institute{Astrophysical Institute
        Potsdam, An der Sternwarte 16, 14482 Potsdam, Germany
       \and
        Universit\"ats-Sternwarte G\"ottingen, Geismarlandstr. 11,
           37083 G\"ottingen, Germany
       }

   \date{Received 14 August 1998 / Accepted 15 September 1998}

   \titlerunning{X-ray properties of V~Sge}
   \authorrunning{J. Greiner \& A. van Teeseling}

   \maketitle

\begin{abstract}
   We investigate the ROSAT X-ray properties of V~Sge, 
   which has been proposed to be related 
   to supersoft X-ray binaries. During optical bright states,
   V~Sge is a faint hard X-ray source, while during optical
   faint states ($V \ga 12$ mag), V~Sge is a `supersoft' X-ray source.
   Spectral fitting confirms that V~Sge's X-ray properties during
   its soft X-ray state may be similar to those of supersoft X-ray
   binaries, although a much lower luminosity cannot be excluded.
   It is possible to explain
   the different optical/X-ray states by a variable amount of
   extended uneclipsed matter,
   which during the optical bright states contributes significantly
   to the optical flux and completely absorbes the soft X-ray component.
   An additional, perhaps permanent, hard X-ray component, such as a
   bremsstrahlung component with a 0.1--2.4\,keV luminosity of
   $\sim 10^{30}$\,erg\,s$^{-1}$, must be present to explain the X-ray
   properties during the optical bright/hard X-ray state.

\end{abstract}

\section{Introduction}

V~Sge is a blue star with a mean brightness around 11 mag which has been shown
to vary between 9.6--14.7 mag since its discovery in 1902. It shows wide 
eclipses at a period of 0\fd51419, a small secondary eclipse, and complex 
emission line behaviour (Herbig \etal\ 1965). Extinction estimates vary between
$E_{\rm B-V}=0.4$ (Herbig \etal\ 1965) and $E_{\rm B-V}=0.15$ 
(Verbunt 1987) implying a distance of 0.7--2.7 kpc.

Supersoft X-ray binaries (SSB; see Greiner 1996 and references therein;
van Teeseling 1998) were established as a new class of accreting
binaries during the early 90ies with ROSAT (Tr\"umper \etal\ 1991;
Greiner \etal\ 1991) and are thought to contain white dwarfs accreting mass at
rates sufficiently high to allow stable nuclear surface burning of the accreted
matter (van den Heuvel \etal\ 1992). SSB have
luminosities of $L_{\rm bol} \sim$\ 10$^{36}$--10$^{38}$ ergs s$^{-1}$, but
their characteristic temperatures of 20--40 eV imply strong attenuation
by the interstellar medium. Thus, most of the known SSB
are located in external galaxies (e.g. Greiner 1996) making detailed optical 
observations difficult. It is therefore of great interest to identify 
galactic SSB.

It has recently been suggested (Steiner \& Diaz 1998; Patterson \etal\ 1998)
that V~Sge has spectroscopic and photometric properties which are very
similar to those of SSB.
This suggestion is based on characteristics which are typical for SSB,
but are rare or even absent among canonical cataclysmic variables: 
(1) the presence of both O\VI\ and N\V\ emission lines,
(2) a He\,\II\,$\lambda 4686$/H$\beta$ emission line ratio $\gax 2$,
(3) rather high absolute magnitudes and very blue colours, and
(4) orbital lightcurves which are characterized by a wide and deep eclipse.

The suggestion of the similarity of V~Sge to SSB 
is almost entirely based on optical and ultraviolet data.
In this paper, we investigate the archival ROSAT data
of V~Sge and discuss them in the context of the long-term optical
behaviour of V~Sge. 
Hoard \etal\ (1996) reported the detection of V~Sge as a soft X-ray source
in the Nov. 1992 ROSAT observation, but did not perform a spectral fit.
Verbunt et al. (1997) already reported the non-detection of V~Sge
during the ROSAT all-sky survey.

\section{ROSAT Observations}

\begin{table*}
\vspace{-0.25cm}
\caption{ROSAT observations of V~Sge}
\vspace{-0.25cm}
\begin{tabular}{clrccccccr}
\hline
\noalign{\smallskip}
 Date & ~Obs-ID$^{(1)}$ & T$_{\rm exp}$~ & offaxis & CR$^{(2)}$ & HR1$^{(3)}$ &
         HR2$^{(3)}$ & X-ray & optical  & D$^{(4)}$ \\
   &  & (sec)~ & angle & (cts/s) & &  &  state & state &  \\
\noalign{\smallskip}
\hline
\noalign{\smallskip}
         Oct. 19--31, 1990 & ~~~-- & ~~~~50 & 0--55\amin & $<$0.054 
                                             & -- & -- & -- & bright & --~~ \\
 Nov. 23/24, 1991 & 400155P & $\!\!\!$10\,235 & 0\farcm39 & 0.0011$\pm$0.0004&
          1.0$\pm$0.7     & 0.2$\pm$0.4   & hard & bright & $\!\!$16\asec \\ 
 Nov. 10--12, 1992 & 300182P & $\!\!\!$27\,745 & 30\farcm9 & 0.0091$\pm$0.0010&
          --0.64$\pm$0.15 & --0.13$\pm$0.39 & soft & intermediate & 28\asec \\
 Apr. 19--24, 1994 & 300311H & $\!\!\!$24\,610 & 0\farcm18 & $<$0.00044 &
          --  &          --             & -- & bright & --~~ \\
 May 12--13, 1994 & 300311H &  4\,700 & 0\farcm18 & 0.0199$\pm$0.0021 & 
          --  &          --            & very soft & faint & 1\asec \\
 Oct. 18/19, 1994 & 300311H-1 & $\!\!\!$18\,440& 0\farcm17 & 0.0011$\pm$0.0003&
          --  &          --            & hard & bright & 5\asec \\
 May 12/13, 1997 & 300582H & $\!\!\!$15\,700 & 0\farcm17 & 0.0025$\pm$0.0004 &
          --  &          --           & soft & intermediate & 1\asec \\
\noalign{\smallskip}
\hline
\end{tabular}

\noindent{\small

 $^{(1)}$  The letter after the observation ID number gives the ROSAT detector:
               P = PSPC, H = HRI. \\
 $^{(2)}$  Count rates in the corresponding detector in the 0.1--2.4 keV range
        (PSPC: channels 11--240). Upper limits are 3$\sigma$ confidence level.
           Note the different PSPC to HRI count rate conversion factors of 
           2.7:1 and 7.8:1 for hard and soft spectrum sources.\\
 $^{(3)}$  Hardness ratios with HR1 = $(B-A)/(B+A)$ and HR2 = $(D-C)/(D+C)$,
	   where $A (0.1-0.4$ keV), $B (0.5-2.0$ keV), $C (0.5-0.9$ keV),
	   and $D (0.9-2.0$ keV) are the counts in the given energy range. \\
 $^{(4)}$ Distance between best-fit X-ray and optical position. For the optical
          position $\alpha$(2000.0)=20\h 20\m 14\fss7, 
           $\delta$(2000.0)=+21\degr 06\amin 10\asec\ has been
          used as determined from the second generation DSS. This position 
          differs from the SIMBAD position by $\triangle\alpha$=12\asec\ and 
          $\triangle\delta$=4\asec. }
\label{log}
\vspace*{-0.1cm}
\end{table*}

   \begin{figure*}
    \vbox{\psfig{figure=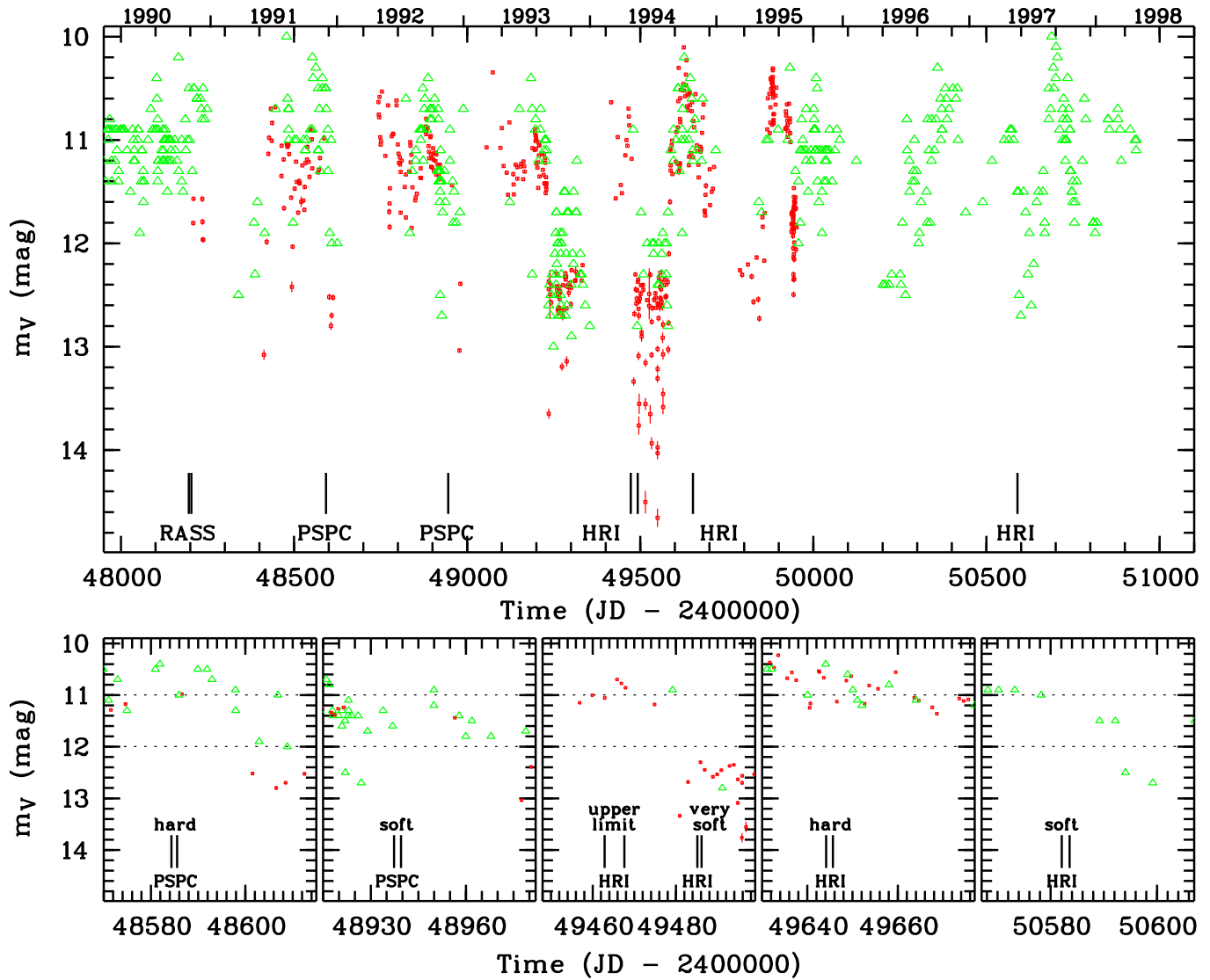,width=18.1cm,%
           bbllx=1.1cm,bblly=1.5cm,bburx=19.1cm,bbury=16.4cm,clip=}}\par
    \vspace*{-0.2cm}
    \caption[lc]{Optical light curve of V~Sge with data from 
            Robertson \etal\ 1997 (red dots) and VSOLJ (Web; green triangles).
            Vertical dashes mark the times of ROSAT observations.
            The lower panels show blow-ups around the ROSAT observations
            which are characterized by two vertical
            lines marking the start and the end of the ROSAT exposure.
            The dotted lines denote the boundaries of the three optical states.
           }
      \label{lc}
   \end{figure*}

V~Sge has been the target of three dedicated pointed PSPC and HRI observations
(one of these splits into 3 separate observation intervals),
and in addition is in the field of view of another PSPC observation
(Table~\ref{log}). The results of these observations are quite diverse:
V~Sge has not been detected during the ROSAT all-sky survey in 1990 and a long
ROSAT HRI pointing in April 1994, but has been detected during all other
observations, even in a much shorter HRI observation. Thus, V~Sge shows
strong X-ray variability with an amplitude of a factor of 140.
In addition, the X-ray spectral characteristics during two ROSAT PSPC
pointings obtained 1 yr apart show a remarkable difference:
at one occasion V~Sge has a `supersoft' X-ray spectrum, at another occasion
the spectrum is very hard.

The diversity of X-ray measurements looks more ordered when it is compared 
with the optical brightness of V~Sge. This binary system is included
in the RoboScope program of automatic long-term monitoring 
the results of which led to the classification of three distinct optical 
states: bright state (V$<$11 mag), intermediate state (V$\sim$11--12 mag)
and faint state (V$>$12 mag)
(Robertson \etal\ 1997).  We have combined the optical lightcurve obtained
by these observations with data collected in the VSOLJ database
(www.kusastro.kyoto-u.ac.jp/vsnet/) and plotted these in
Fig.~\ref{lc} together with the times of the ROSAT observations.
This suggests that during optical bright state V~Sge is 
a hard, but rather faint X-ray source, while during optical faint state 
V~Sge is a more luminous and very soft X-ray source. 
During the intermediate
optical state also the X-ray spectrum is intermediate with respect to the
very soft and hard spectrum.

\begin{figure}
    \vbox{\psfig{figure=vsge_logg9.ps,width=8.8cm,angle=270}}
    \vspace{0.3cm}
    \vbox{\psfig{figure=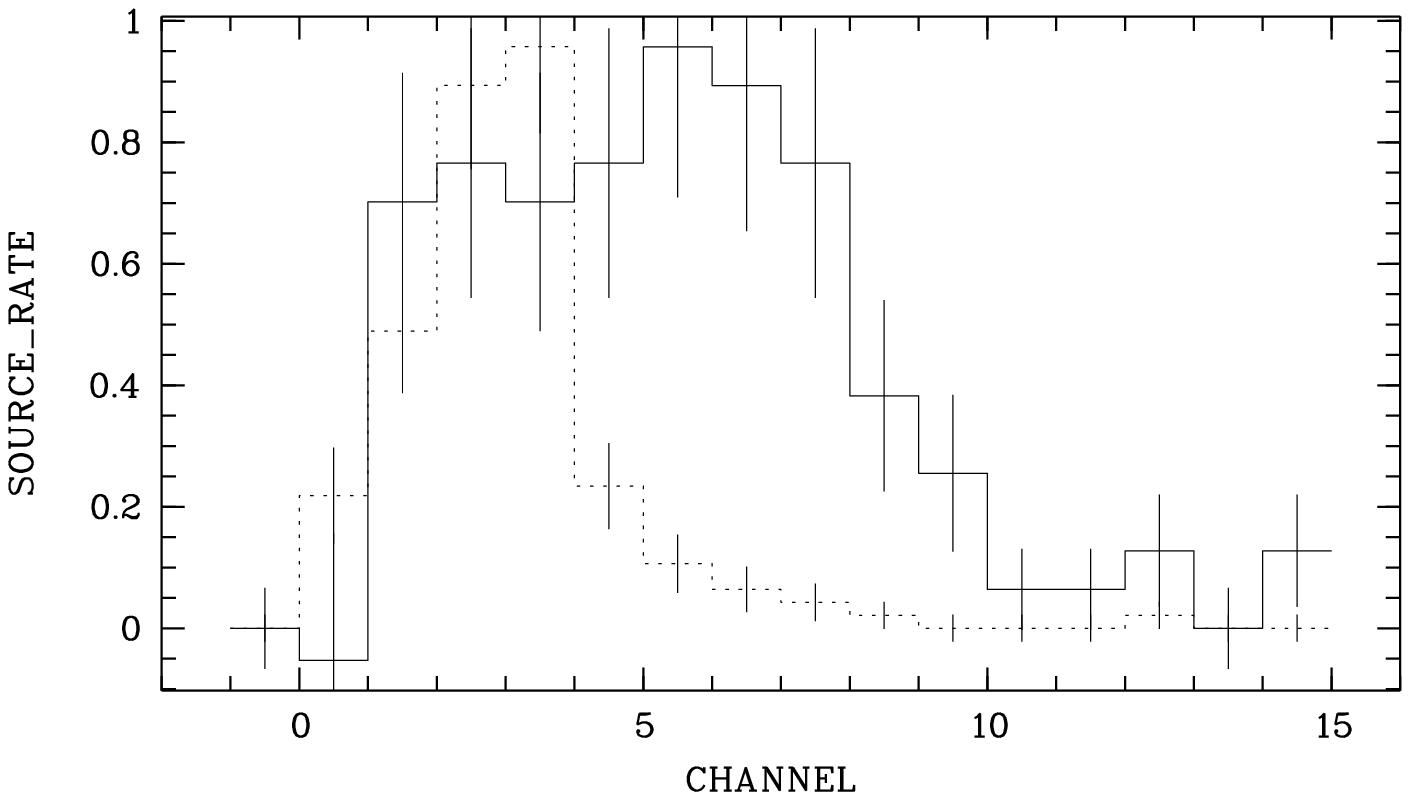,width=8.8cm,%
            bbllx=2.8cm,bblly=4.cm,bburx=16.3cm,bbury=12.1cm,clip=}}\par
    \caption[logg]{{\bf Top:} $1\sigma$ and 90\% confidence contours (solid 
           lines) of a $\chi^2$ fit of solar-abundance LTE $\log g = 9$
           model spectra to the Nov. 1992 ROSAT PSPC spectrum,
           when V~Sge was in an intermediate optical state and had
	   a very soft X-ray spectrum.
           Dashed lines denote contours of constant absorbing column 
           $n_{\rm H}$/$10^{20}$\,cm$^{-2}$,
           dotted lines indicate white dwarf radii in units of $10^8$\,cm. 
	   The radius and luminosity have been scaled to a distance of 1\,kpc.
           {\bf Bottom:} Comparison of the normalized HRI channel distribution
           of the photons during the May 1994 (dotted line; soft state) and
           Oct. 1994 (solid line; hard state) observations.
          }
      \label{logg9}
   \end{figure}

To obtain an idea about the X-ray spectral parameters during the
soft X-ray state, we fit the Nov. 1992 PSPC spectrum with a solar-abundance
LTE $\log g = 9$ white dwarf atmosphere model (Van Teeseling \etal\ 1994).
The $\chi^2$ contours are shown in Fig.~\ref{logg9}. 
The $1\sigma$ contour suggests a temperature
$T_{\rm eff} > 500\,000$\,K and a bolometric luminosity
$L \lax 10^{33}$\,erg\,s$^{-1}$ (with $d = 1$\,kpc), but lower temperatures
and higher luminosities are still acceptable within the 90\% confidence 
contour.
If we require that the soft X-ray absorbing column is at least
$n_{\rm H}\sim 8\times 10^{20}$\,cm$^{-2}$ as derived from the $2200$\,\AA\
absorption dip ($E(B-V)\sim 0.15$; Verbunt 1987), we find 
$T_{\rm eff} < 800\,000$\,K and $L > 10^{32}$\,erg\,s$^{-1}$.
With $n_{\rm H}\sim 10^{21}$\,cm$^{-2}$, only for $T_{\rm eff}\lax$200\,000\,K
a luminosity of $L > 10^{36}$\,erg\,s$^{-1}$ is reached.
It is possible, however, that because of the very high orbital inclination
the soft X-ray absorption is much larger than the ultraviolet absorption.
A similar discrepancy is known for CAL\,87 (cf. Hutchings \etal\ 1995;
Parmar \etal\ 1997). If we relax the absorption constraint and assume that 
V~Sge is a SSB
with $L > 10^{36}$\,erg\,s$^{-1}$, we find $T_{\rm eff} < 500\,000$\,K
and a radius $R > 2\times 10^8$\,cm consistent with a white dwarf.
However, because of a very high orbital inclination, the white dwarf
may be completely obscured from view by the accretion disk rim, in
which case the observable luminosity (from X-rays scattered into the
line of sight) may be much less than $10^{36}$\,erg\,s$^{-1}$.
We note that there is no significant modulation of the soft X-rays on the
orbital period.

A factor of 45 increase in HRI count rate occurred within less than three 
weeks in April/May 1994 during
which the optical brightness decreased and V~Sge eventually became a very
soft X-ray source. Though the ROSAT observations during this optical state
transition have been performed with the HRI, the grossly different spectral 
shapes are easy to recognize (Fig. \ref{logg9}, bottom).

\section{Discussion}

The anti-correlation of soft X-ray emission with optical brightness
is reminiscent of the behaviour of the SSB \0513\
(Reinsch \etal\ 1996; Southwell \etal\ 1996). \0513\ turns on as a
supersoft X-ray source only during $\sim 1$\,mag optical dips, which
occur every 100--200 days and last about $\sim 30$ days. This
behaviour has been explained by assuming that the shell-burning
white dwarf in \0513\ has normally expanded to a few $10^{10}$\,cm
and radiates its luminosity in the extreme-ultraviolet. During the
optical faint states, the white dwarf contracts with almost constant
bolometric luminosity to $\sim 10^9$\,cm, and radiates 
in the soft X-ray band.

The model that has been suggested for \0513\
cannot explain the observational data of V~Sge. First, the optical
brightness changes of V~Sge are very rapid: both the faint-/bright-state
transitions as well as the succession of different faint states may occur 
on timescales of $\lax 1$\,day (compared to the smooth decline
of several days in \0513). Such very rapid changes are only possible
if the white dwarf envelope expands and contracts on the Kelvin-Helmholtz
timescale and the mass of the expanding envelope is rather small 
($M_{\rm env}\sim 10^{-9} M_{\sun}$). Such a small envelope mass
is difficult to accept for a white dwarf with stable shell burning
(e.g. Prialnik \& Kovetz 1995). Second, the expected
optical eclipse would become deeper when the system becomes brighter,
opposite to what has been observed (Patterson \etal\ 1998).

Before we speculate on a possible explanation for the observed X-ray 
properties of V~Sge, we note the similarity of the V~Sge behaviour to 
that of VY~Scl stars (as has been noted with respect to the
optical behaviour also by Robertson \etal\ 1997). In a recent survey of the
available ROSAT data of VY~Scl stars (Greiner 1998) a relatively hard
X-ray spectrum was found for VY Scl stars during optical bright state
(see also van Teeseling \etal\ 1996).
Moreover, observations of the VY~Scl star V751~Cyg during its 1997 optical
faint state have revealed luminous and very soft X-ray emission 
(Greiner \etal\ 1998), similar to that of V~Sge in its faint state.

Inspection of the change in eclipse depth from faint to bright state 
(e.g. Fig.~5 in Patterson \etal\ 1998) shows that it is possible to
reproduce this change by an increase of uneclipsed flux,
while the eclipsed light (presumably from the irradiated accretion disk)
remains almost constant. If the flux from the irradiated disk (and
therefore also from the irradiated secondary) remains unchanged, this
would suggest that a brightening of V~Sge is caused by an increasing amount
of extended luminous (outflowing?) matter.
This would also be consistent with the
emission lines, which only show partial eclipse effects, and
which increase in strength when V~Sge brightens
(e.g. Herbig \etal\ 1965), indicating either an increasing amount
of line emitting matter or an increasing amount of ionizing flux.
(Mauche \etal\ (1997) note that the
He\,{\sc ii}\,$\lambda 1640$/C\,{\sc iv}\,$\lambda 1550$ and
N\,{\sc v}\,$\lambda 1240$/C\,{\sc iv}\,$\lambda 1550$ line ratios
decrease when V~Sge brightens.)

Additional extended gas could increase the amount of soft X-ray
absorption and make the soft X-ray component fainter and harder
and make it even completely undetectable. Because it is impossible to
produce the hardness ratios {\em and} count rate observed in
Nov. 1991 by simply adding more absorption to an absorbed hot-white-dwarf
spectrum, we conclude that in Nov. 1991 an additional
harder X-ray component was present. The hardness ratios and count rate
of the Nov. 1991 PSPC observation can be explained perfectly with a
thermal bremsstrahlung spectrum, absorbed with
$n_{\rm H}\gax 10^{21}$\,cm$^{-2}$, a temperature of a few keV,
and a 0.1--2.4\,keV luminosity of $\sim 10^{30}$\,erg\,s$^{-1}$ (at 1\,kpc).
The same bremsstrahlung component may also have been present during the
soft X-ray state in Nov. 1992 without significantly affecting the
confidence intervals in Fig.~\ref{logg9} of the spectral parameters of the
soft component.
We note that such a bremsstrahlung component is not inconsistent with the
X-ray flux of an evolved secondary in a 12\,hr binary (cf. Dempsey \etal\
1993).

A simple wind model for the recently observed radio flux density of V~Sge
implies a mass-loss rate of the order of 10$^{-6}$\,\msun/yr
(Lockley \etal\ 1997). With their (assumed) terminal velocity of 1500 km/s 
this wind zone is completely opaque for X-rays up to 0.7\,keV, even if the
wind is assumed to be circumbinary instead of arising from one component.
Since the radio measurement has been obtained during optical high state,
it supports the above described scenario. We note that the colliding wind
scenario as discussed in Lockley \etal\ (1997) would predict a positive 
correlation between optical and X-ray emission, contrary to our finding.
Vitello \& Shlosman (1993) have modeled the UV line shapes of V~Sge assuming a
biconical accretion disk wind, and need high mass-loss rates to
explain the observations, while they did not consider the
possibility of a luminous shell-burning white dwarf.
We also note that irradiation-induced winds with a rate of
10$^{-7}$--10$^{-6}$\,\msun/yr are expected in SSB 
(Van Teeseling \& King 1998).

We conclude that it is possible to explain the optical and X-ray
behaviour of V~Sge by assuming a variable amount of extended
uneclipsed matter which contributes significantly in the optical
(by reprocessing of soft X-rays?) and may completely absorb the
`supersoft' X-ray component during the optical bright/hard X-ray state.
The X-ray properties of V Sge, in any case, support the presence
of a hot luminous white dwarf, possibly with hydrogen shell-burning, and
do not hinder the addition of V Sge to the class of supersoft X-ray binaries.
More detailed modelling of the changing optical light curves
and ultraviolet and optical spectra are necessary to test this scenario.

\begin{acknowledgements}
We thank Kent Honeycutt for providing his RoboScope data as published
in Robertson \etal\ (1997) in electronic format. 
JG and AvT are supported by the German Bundesmi\-ni\-sterium f\"ur Bildung,
Wissenschaft, Forschung und Technologie
(BMBF/DLR) under contract No. FKZ 50 QQ 9602 3 and 50\,OR\,96\,09\,8,
respectively.
The \ros\, project is supported by BMBF/DLR and the Max-Planck-Society.
This research has made use of the Simbad database, operated at CDS, 
Strasbourg, France and the Digitized Sky Survey (DSS) produced at 
the Space Telescope Science Institute under US Government grant NAG W-2166.
\end{acknowledgements}

\end{document}